\documentstyle[12pt]{article}

\begin{document}

{\pagestyle{empty} }

\vskip 6mm

\centerline{\large \bf
Scalar Field Contribution to 
 Rotating Black Hole Entropy }

\vspace{10mm}

\centerline{Masakatsu Kenmoku \footnote{E-mail address:
        kenmoku@asuka.phys.nara-wu.ac.jp} , 
        Kimiko Ishimoto \footnote{E-mail address:
        ishimoto@asuka.phys.nara-wu.ac.jp}}
\centerline{\it Department of Physics,
Nara Women's University, Nara 630-8506, Japan}

\vskip 3mm
\centerline{ Kamal Kanti Nandi \footnote{E-mail address:
              kamalnandi1952@yahoo.co.in }}
\centerline{\it Department of Mathematics,
University of North Bengal}
\centerline {\it Darjeeling (W.B.) 734 430,
India }

\vskip 3mm
\centerline{Kazuyasu Shigemoto \footnote{E-mail address:
        shigemot@tezukayama-u.ac.jp} }
\centerline{\it Department of Physics,
Tezukayama University, Nara 630-8501, Japan} 

\vskip 1cm \centerline{\bf Abstract}

Scalar field contribution to entropy is
studied in arbitrary $D$ dimensional one parameter 
rotating spacetime by semiclassical
method. By introducing the zenithal angle dependent cutoff parameter, the
generalized area law is derived. The non-rotating limit can be taken
smoothly and it yields known results. The derived area law is then applied
to the Ba\~{n}ados-Teitelboim-Zanelli (BTZ) black hole in (2+1) dimension
and the Kerr-Newman black hole in (3+1) dimension. The generalized area law 
is reconfirmed by the Euclidean path integral method for the quantized scalar
field. The scalar field mass contribution is discussed briefly.

\vskip 3mm

\noindent PACS number(s): 04.62.+v, 04.70.-s, 04.70.Dy, 
97.60.Lf \hfil \vfill

\newpage

\setcounter{equation}{0}
\renewcommand{\theequation}{\thesection.\arabic{equation}}

\section{Introduction}
One of the important offshoots of general relativity is black hole physics.
Extensive investigations in this field have thrown up  riches of
information that are significant from the theoretical and/or observational
point of view \cite{mtw:,townsend:}. On the theoretical side, many exact
black hole solutions in general relativity are 
well known \cite{stephani:,mp:}.
Among them, stationary, axi-symmetric exact solution with asymptotic flatness
is known as the Kerr-Newman solution that represents charged, massive,
rotating black hole in (3+1) dimensional spacetime \cite{kerr:,newman:}.
Stationary rotating black hole solution with a negative cosmological term is
known as the BTZ solution which asymptotically approaches the anti-de Sitter
spacetime in (2+1) dimension \cite{btz:}. 
Recently, Kerr-AdS black hole solutions in higher dimensions 
is studied with the interest of AdS/CFT correspondence 
\cite{h:,g:,y:}. 
Black holes are assumed to attract
many matter fields due to their strong gravitational force but they scatter
out almost all the matter information in the form of gravitational or scalar
wave radiation so that the final stage is characterized by only three hairs:
their mass, angular momentum and charge \cite{ruffine:}. The rotation
effects induce the ergo region wherein no particle can remain at rest. The
ergo region extends to the outsude from the horizon, and in this region
negative energy of particles can be absorbed into black holes if the particles
have angular velocities in the direction opposite to that of the black hole.
This effect induces the so called superradiance phenomenon \cite{thorne:},
namely, that the energy of the scattered particle can be larger than that of
the incident particle, and is generally termed as the Penrose process 
\cite{penrose:,cheist:}.

Black holes may interact strongly with matter fields and are thought of as
thermal objects. From the analogy between the laws of thermodynamics and the
laws of black holes, the area law of black holes has been proposed. 
(A recent article by Bekenstein \cite{bekenstein0:} is also recommended.)
Since then, this law (Entropy $\propto $ Area) has played a pivotal role
in the understanding of black hole physics in general 
\cite{bekenstein:,bardeen:,hawking:,gour:}. 
A remarkable consequence of the proportionality is its
connection with the holographic principle which produces Einstein's
equations and other thermodynamical relations
\cite{jacobson:}. 
The relation of the holographic entropy bound
with QFT is investigated by Yurtsever
\cite{yurt:}. 
The area law can be proved by many methods and approximations using
the classical black hole solutions of general relativity. The matter field
contribution to the black hole entropy has also been studied extensively
using some of the methods \cite{hartle:,dealwis:,thooft:,susskind:}. Among
them, the semiclassical method with the brick wall regularization scheme
seems to be more transparent from the mathematical and physical points of
view \cite{thooft:}. These features become evident when the method is
applied especially to the well known cases of static Schwarzschild and
Reissner-Nordstr\"{o}m black holes\cite{kenmoku:}. 

However, the statistical mechanics for the matter field contribution in the
background of a rotating black hole is somewhat problematic. 
Calculations involving rotating black holes are relatively few and diverse.
Some examples
are as follows: The entropy of the Kerr-Newman black hole in (3+1) dimension
has been studied by several authors \cite{ho:,lee:,mann:,kim:} and the extra
divergent structures were pointed out to appear due to the superradiant
mode. More recent works are found in \cite{recent:}. A somewhat early work using
a complex metric and Jacobi action may be found in \cite{brown:}.
The area law of a rotating black hole emerged only if some additional
cutoff parameters were introduced and a special relation between them were
imposed. The non-rotating limit of the entropy cannot be taken in these
calculations. Likewise, the entropy of BTZ black hole in (2+1) dimension has
been studied by several authors \cite{ichinose:,kimkim:,fatibene:,ho2:} . 
In this case too, the
extra divergences appeared due to the superradiant mode and in order to
evade these divergences, extra cutoff parameters were introduced. The area
law has been derived by fixing the cutoff parameter artificially. Again, the
area law does not yield a smooth non-rotating limit. Moreover, the free
energy and the entropy from the scalar field contributionin (2+1) dimensional
black hole spacetime was reported to depend largely on the specific
approach adopted to calculate them \cite{ichinose:}. The problems outlined
here revealed themselves in the analyses dealing with specific solutions. It
is therefore desirable that the calculations be carried out in a
sufficiently general framework such that the above difficulties are either
removed or at least minimized as far as possible. A solution independent
generalized area law is expected to provide a better physical insight in the
understanding of rotating black hole thermodynamics.

In this paper, we calculate the scalar field contribution to the black hole
entropy in an arbitrary $D$ dimensional rotating black hole spacetime {\it %
without} assuming any particular exact solution. Higher dimensional
calculations of this kind may be useful in string theory and M-theory as
well \cite{youm:}. 
We restrict to treat one parameter rotating black holes in this paper, 
which includes many essential aspects \cite{k:}.    
The analysis here is
carried out under a minimal set of physical requirements:\ The generic
metric components are assumed to be independent of time $t$ and azimuthal
angle $\phi $ so that the energy and angular momentum are conserved. The
off-diagonal metric component $g_{t\phi }$ is assumed to exist which
indicates the rotation of the black hole. We shall first adopt the semiclassical
method for the statistical mechanics of the scalar field and derive the
generalized Stefan-Boltzmann's law with the help of near horizon
approximation. In addition, we shall impose the horizen and the 
temperature conditions on the time
and radial components of the metric, restricting to the black hole with 
simple zero at the horizen and making the temperature well-defined.
Under these conditions, we derive the
generalized area law of rotating black holes in $D$ dimension. In the
derivation, we introduce the zenithal angle dependent cutoff parameter 
$\epsilon (\theta )$ consistent with the brick wall regularization
scheme. 
We then apply the result to the known rotating black hole solutions: BTZ and
Kerr-Newman. Superradiant modes are included and the non-rotating limit can
be taken smoothly in our calculation. We shall next adopt the Euclidean path
integral method to reconfirm the generalized area law.

This paper is organized as follows. The generalized Stefan-Boltzmann's law for
the massless scalar field will be derived via semiclassical method in
section 2.1. The generalized area law will be derived using the brick
wall regularization scheme in section 2.2. This law is then applied to (2+1)
and (3+1) dimensional black holes in section 3. The area law is
reconfirmed by the Euclidean path integral method in section 4. The small
scalar field mass contribution to the free energy will be discussed 
at the end of section 4. 
In the final section, the results will be summarized and will 
make some some discussions.

\section{Scalar Field Contribution to Rotating Black Hole Entropy 
by Semiclassical Method  }

In this section, we study the statistical mechanics for the scalar field in 
$D$ dimensional one parameter rotating spacetime by the semiclassical method. 
Our analysis is general in the sense that that the metric does not
depend on the explicit black hole solution. 
For the
massless scalar case, we obtain the generalized form of the Stefan-Boltzmann's
law for the free energy, entropy and the internal energy near the rotating
black hole horizon. As the real space integral diverges due to the large
time delay near the horizon, we introduce the short distance cutoff
parameter in the brick wall regularization scheme developed by 't Hooft 
\cite {thooft:}. We can then derive the generalized area law for the rotating
black holes using the zenithal angle dependent cutoff parameter. 
We adopt units such that 
$c=\hbar=k_{{\rm B}}=1$ unless otherwise specified.

\subsection{Stephan-Boltzmann's law 
in rotating black hole spacetime }

In order to study the statistical properties of the 
black hole entropy in one rotation parameter case,  
we apply the semiclassical method (or WKB method) and 
set the $D$ dimensional polar coordinate as \begin{eqnarray}
x^{\rho}=(x^0, x^1, x^2, \cdots , x^{D-1} )
=(t, r, \phi, \theta^{3}, \cdots , \theta^{D-1})
\ \ , \label{e1}
\end{eqnarray}
where the ranges are  
$t\in (-\infty,\infty)$, $r\in[0,\infty)$, $\phi\in[0,2\pi]$ and  
$\theta^{3}, \cdots ,\theta^{D-1}\in[0,\pi] $. 
The invariant line element is assumed to be of the form
\begin{eqnarray}
ds^2=g_{tt}dt^2+g_{rr}dr^2+g_{\phi \phi}{d\phi}^2
+2g_{t\phi}dtd\phi +\sum_{i=3}^{D-1}g_{ii}(d\theta^{i})^2\ , \label{e2}
\end{eqnarray}
where the off diagonal metric $g_{t\phi}$ induces the rotation 
of the system with the angular velocity $-g_{t\phi }/g_{\phi \phi }$.   
The metric components in Eq.(\ref{e2}) do not 
depend on $t, \phi$ and can 
depend on $r$ and $\theta^{i}\ (i=3, \cdots , D-1)$. 
Consequently, two Killing vectors exist:    
\begin{eqnarray}
\xi_{t}^{\rho}=(1,0,\cdots,0)\ , \ \ 
\xi_{\phi}^{\rho}=(0, 0, 1, 0, \cdots, 0)\ , 
\label{e3}
\end{eqnarray}
which imply the conservations of 
the total energy $E$ and the azimuthal angular momentum $m$ 
of a scalar field.  


With the metric components in Eq.(\ref{e2}), 
the matter action for the scalar field $\Phi$ of mass $\mu $ 
in $D$ dimension is
\begin{eqnarray}
I_{\rm matter}(\Phi )
=\int d^{D}x\sqrt{-g}\left(
-\frac{1}{2} g^{\rho \sigma }\ \partial_{\rho}\Phi 
\partial _{\sigma}\Phi 
-\frac{1}{2} \mu^{2}\Phi^2 \right) \ .
\label{e4}
\end{eqnarray}
From this action,  
the field equation for the scalar field is obtained:  
\begin{eqnarray}
\frac{1}{\sqrt{-g}}\partial _{\rho}
\left( \ \sqrt{-g}g^{\rho \sigma }\partial
_{\sigma}\Phi \right) -\mu^{2}\Phi =0\ .
\label{e5}
\end{eqnarray}
We take the ansatz for the scalar field $\Phi $ 
in the semiclassical method as 
\begin{eqnarray}
\Phi (t, r, \phi, \theta^{3}, \cdots , \theta^{D-1})
\simeq\exp 
\left(i \sum_{\rho=0}^{D-1} \int_{Q_0}^{Q}
 p_{\rho}dx^{\rho}\right) \ ,  
\label{e6}
\end{eqnarray}
where 
the line integral is performed from the fixed point $Q_{0}$ 
to an end point $Q=(t,\phi,r,\theta^{3},...,\theta^{D-1}, r)$ and  
$p_{\rho}$ denotes the $D$ dimensional momentum whose components 
are 
\begin{eqnarray}
p_{\rho}=(-E, p_r, m, p_{3}, \cdots , p_{D-1}) \ .
\label{e7}
\end{eqnarray}
Putting the scalar field function Eq.(\ref{e6}) into 
the field equation Eq.(\ref{e5}) and ignoring 
the derivatives of the metric components, 
the on-shell energy-momentum relation is obtained 
\begin{eqnarray}
-\mu^2
=g^{\rho\sigma}p_{\rho}p_{\sigma}
=g^{tt}E^2+g^{rr} p_r^2 +g^{\phi\phi}p_{\phi}^2-2g^{t\phi} E m
+\sum_{i=3}^{D-1}g^{ii}p_{i}^2\ ,
\label{e8}
\end{eqnarray}
where the contravariant components of the metric are obtained from
the original metric Eq.(\ref{e2}) as 
\begin{eqnarray}
&&g^{tt}=g_{\phi\phi}/\Gamma \ \ , \ \ 
g^{rr}=1/g_{rr} \ \ , \ \ 
g^{\phi\phi}=g_{tt}/\Gamma \ \ , \ \ 
\nonumber \\
&&g^{t\phi}=-g_{t\phi}/\Gamma \ \ , \ \ 
g^{ii}=1/g_{ii} \ \ (i=3, \cdots , D-1) \ \ , \ \ 
\label{e9}\ 
\end{eqnarray}
with $\Gamma:=g_{tt}g_{\phi\phi}-g_{t\phi}^2$.

We note that using the action of a particle picture of mass $\mu$ 
in rotating spacetime:    
\begin{eqnarray}
 I_{\rm particle}=-\mu\int dt
\sqrt{-
g^{\rho\sigma} \frac{dx^{\rho}}{dt}
               \frac{dx^{\sigma}}{dt}
} \ , \label{e10}
\end{eqnarray}
the energy conservation, 
the angular momentum conservation  
and the mass shell condition of the particle 
corresponding to Eq.(\ref{e8}) can be derived.

Since the black hole is rotating, its energy can be transferred 
to scalar particles in the ergo region  
by the Penrose process \cite{penrose:}. However, not all the 
black hole energy can be mined out. 
We can obtain the restriction on energy $E$ 
and angular momentum $m$ of the scalar particle 
in the following way \cite{cheist:}. 
Consider a new Killing vector combining 
two Killing vectors in Eq.(\ref{e3}) linearly as 
\begin{eqnarray}
\eta:=\xi_{t}+\xi_{\phi}\Omega_{H} \ , \label{e11}  
\end{eqnarray}
where the angular velocity on the horizon $r_{H}$ is defined as
\begin{eqnarray}
\Omega_{H}:=\left. \frac{g^{t\phi}}{g^{tt}}\right|_{r_{H}}
=-\left. \frac{g_{t\phi}}{g_{\phi\phi}}\right|_{r_{H}}
\ . \label{e12}
\end{eqnarray}
This vector is light like in future direction 
on the horizon 
\footnote[1]{The light like property of vector $\eta$ is shown using the 
horizon condition in subsection 2.2: $1/g^{tt}=0$ on the horizon.} 
and the inner product of 
it with momentum in Eq.(\ref{e7}) becomes non-positive, 
which provides the restriction on the energy: 
\begin{eqnarray} 
p\cdot \eta:=
\sum_{\rho=0}^{D-1} p_{\rho} \eta^{\rho} \leq 0 \ \ \Rightarrow \ \ 
m \Omega_{H} \leq E  \ .  
\label{e13}
\end{eqnarray}
This means that 
the angular momentum of scalar particle is inverse in sign 
to the angular velocity of the black hole  
if the negative energy particle is absorbed into the black hole.  

The semiclassical quantization condition is imposed to 
require the  single valueness for the scalar wave function Eq.(\ref{e6}) 
in $(D-1)$ dimensional space, 
and the number of the quantum state with energy not exceed $E$ is 
the sum of phase space $K$ divided by the unit quantum volume such that
\begin{eqnarray}
\sum_{K} \simeq \frac{1}{2\pi^{D-1}}
\int dr dp_r d\phi dm  d\theta^{3} dp_{3} \cdots
d\theta^{D-1} dp_{D-1}\ , \label{e14} 
\end{eqnarray}
where the integration range is the range shown 
below Eq.(\ref{e1}) for angle variables and 
the range satisfying the energy-momentum condition Eq.(\ref{e8}) 
with the restriction Eq.(\ref{e13}) for the momentum variables 
near the horizon region. 


Next we consider the partition function of the scalar field 
in the rotating black hole geometry, viz.,   
\begin{eqnarray}
Z&=&\sum_{\{n^{(K)}\}} 
\exp\left( {-\beta \sum_{\{K\}}n^{(K)}(E_{(K)}-m\Omega_{H})} \right) 
\nonumber \\ 
&=& \prod_{\{K\}}\sum_{\{n^{(K)}\}} 
\exp{\left(-\beta n^{(K)}(E_{(K)}-m\Omega_H)\right)} 
\nonumber \\
&=& \prod_{\{K\}}
\left[1-\exp\left( -\beta(E_{(K)}-m\Omega_{H}
)\right)\right]^{-1}
\label{e15} \ ,
\end{eqnarray}
where $\beta$ denotes the inverse temperature. 
The exponent of the Boltzmann factor $E-m\Omega_{H}$ is understood 
taking account of the rotation effect according to the 
Hartle-Hawking argument \cite{hartle:} 
and is positive as shown in Eq.(\ref{e13}) 
ensuring that the partition function is well defined. 
The summation with respect to $n^{(K)}$ and $K$ are 
the occupation number sum and 
the phase space sum in Eq.(\ref{e15}) respectively.

The free energy $F$ 
is obtained through the partition function :
\begin{eqnarray}
\beta F=-\log Z 
= \sum_{\{K\}} \log \left[1-\exp\left( -\beta(E_{(K)}-m\Omega_{H}
)\right)\right]
\label{e16}\ .
\end{eqnarray}
Using the semiclassical phase space sum in Eq.(\ref{e14}) 
and changing the integration variable from $p_{r}$ to $E$, 
we obtain the expression of the free energy after the integration 
by parts in the form 
\begin{eqnarray}
F =&-&\frac{1}{2\pi^{D-1}}\int_{m \Omega_{H}}^{\infty}dE
  \int d\phi dm 
d\theta^{3} dp_{3} \cdots d\theta^{D-1} dp_{D-1} \nonumber \\ 
&\times& \, \int dr \, \frac{p_{r}}{{\rm e}^{\beta(E-m \Omega_{H})}-1} \ , 
\label{e17}
\end{eqnarray}
where the radial momentum $p_{r}$ is determined by 
the mass-shell energy-momentum condition 
Eq.(\ref{e8}) as 
\begin{eqnarray}
p_{r}&=&\frac{1}{(g^{rr})^{1/2}}
\left(-g^{tt}E^2-g^{\phi\phi} m^2+2g^{t\phi}E m
-\sum_{i=3}^{D-1} g^{ii} p_{i}^2-\mu^2 \right)^{1/2}  \nonumber  \\
&=&(g_{rr})^{1/2}
\left(-g^{tt}(E+\frac{g_{t\phi}}{g_{\phi\phi}}m)^2
-\frac{m^2}{g_{\phi\phi}}
-\sum_{i=3}^{D-1}\frac{p_{i}^2}{g_{ii}}-\mu^2 \right)^{1/2} 
.  \label{e18}
\end{eqnarray}
The inverse metric relations in Eq.(\ref{e9}) have been used 
in the second equality. 
\footnote[2]{Note that the inverse metric component $g^{tt}$, 
$g^{\phi\phi}$  respectively is not equal to $1/g_{tt}$, $1/g_{\phi\phi}$ 
in rotating geometry in general. }

In view of the situation of rotating geometry, 
we introduce a new energy variable as     
\begin{eqnarray}
E':=-p\cdot\eta=E-m \Omega_{H} \ .  
\label{e19}
\end{eqnarray}
We now make the near horizon approximation 
on the angular velocity as 
\begin{eqnarray}
-g_{t\phi}/g_{\phi\phi}\simeq\Omega_{H} \ \ 
\mbox{for} \ \  r\simeq r_{H} \ ,  \label{e20}
\end{eqnarray}
so that the radial momentum becomes 
\begin{eqnarray}
p_{r}&\simeq&{(g_{rr})^{1/2}}
\left(-g^{tt}E'^{\, 2}-\frac{m^2}{g_{\phi\phi}}
-\sum_{i=1}^{D-1}\frac{p_{i}^2}{g_{ii}}-\mu^2 \right)^{1/2} 
. \label{e20.1} 
\end{eqnarray}
The energy and momentum variables are changed to dimensionless ones: 
\begin{eqnarray}
x&=&\beta E' \ \ , \ \ y_{2}=Y\frac{m}{\sqrt{g_{\phi\phi}}} 
\ , \nonumber \\    
y_{3}&=&Y\frac{p_{3}}{\sqrt{g_{33}}}\ , \ \cdots \ , \ 
y_{D-1}=Y\frac{p_{D-1}}{\sqrt{g_{{D-1} {D-1}}} }
\ ,   \label{e21}
\end{eqnarray}
with 
\begin{eqnarray}
Y=(-g^{tt}E'^2 - \mu^2)^{-{1/2}}\ . \nonumber 
\end{eqnarray}
Then the free energy is expressed as
\begin{eqnarray}
F&=&-\frac{1}{2\pi^{D-1}\beta^{D}}\int_{0}^{\infty}dx
\frac{x^{D-1}}{{\rm e}^x-1} 
\int dr \int d\phi d\theta^{3} \cdots d\theta^{D-1} 
(-g^{tt})^{D/2-1/2}
\nonumber \\
&& \times (g_{rr} g_{\phi\phi} g_{33} \cdots g_{{D-1} {D-1}})^{1/2}
\left(1+\frac{\beta^2 \mu^2}{g^{tt}x^2}\right)^{D/2-1/2} 
\frac{v_{\rm unit}}{2^{D-2}} 
\ , \label{e22}
\end{eqnarray}
where the term $v_{\rm unit}$ denotes the $(D-1)$ dimensional volume 
of unit sphere 
from the dimensionless momentum $y_{i}$-integration:
\begin{eqnarray}
&&{v_{\rm unit}}
\nonumber \\
&&={2^{D-2}} 
\int_{-1}^{1}dy_2 \int_0^{\sqrt{1-y_{2}^2}} dy_3 \cdots 
\int_0^{\sqrt{1-y_{2}^2- \cdots-y_{D-1}^2}} dy_{D-1} 
\sqrt{1-y_2^2-\cdots y_{D-1}^2} \nonumber \\
&&=\frac{\pi^{D/2-1/2}}{\Gamma(D/2+1/2)}
={2^{D-1}}{\pi^{D/2-1}}\frac{\Gamma(D/2)}{\Gamma(D)}
\ . \label{e23}  
\end{eqnarray}

In the following, we consider the massless scalar field case 
$(\mu=0)$, which 
corresponds to the high temperature case. 
The mass contribution to the free energy will be discussed 
in the last part of section 4.  
The free energy for the massless case is obtained in a compact form 
\begin{eqnarray}
F=-\frac{\zeta(D)\Gamma(D)}{(2\pi)^{D-1}\beta^{D}}
{v_{\rm unit}}{V_{\rm opt}}
\ , \label{e24}
\end{eqnarray}
where the dimensionless energy $x$-integration is carried out 
using the formula 
\begin{eqnarray}
\int_{0}^{\infty}dx\frac{x^{D-1}}{{\rm e}^{ x}-1}
=\zeta{(D)}\Gamma(D) \ , \label{e25}
\end{eqnarray}
and the optical volume is defined as
\begin{eqnarray}
V_{\rm opt}:=\int dr d\phi d\theta^{3}...d\theta^{D-1}
(-g^{tt})^{D/2-1/2}
(g_{rr} g_{\phi\phi}g_{33} \cdots g_{{D-1} {D-1}} )^{1/2}
\ . \label{e26}
\end{eqnarray}
The entropy and the internal energy are then given by 
\begin{eqnarray}
S&:=&\beta^2\frac{\partial F}{\partial \beta} 
=\frac{\zeta(D)\Gamma(D+1)}{(2\pi\beta)^{D-1}}
{v_{\rm unit}}V_{\rm opt} 
\ , \label{e27}\\
U&:=&F+S/\beta =\frac{(D-1)\zeta(D)\Gamma(D)}{(2\pi)^{D-1}\beta^{D}}
{v_{\rm unit}}V_{\rm opt} 
\ .\label{e28}
\end{eqnarray}
The rotation effects are included in $V_{\rm opt}$ and may be 
in $\beta$. 

In order to confirm these thermodynamical formulae, we calculate 
the internal energy in $D=4$ flat spacetime:   
\begin{eqnarray}
U_{D=4}=\frac{1}{2}\times \frac{\pi^2}{15\beta^4}V_{\rm opt}
\ , \label{e29}
\end{eqnarray}
which agrees with the Stefan-Boltzmann's law up to a
numerical factor 
if we take into account the photon polarization freedom 2. 
Therefore, thermodynamical formulae in Eqs.(\ref{e24}), 
(\ref{e27}) and (\ref{e28}) are recognized as 
the generalization of the Stefan-Boltzmann's law to that  
in the $D$ dimensional rotating spacetime. 
 
Note also that the non-rotating limit can be taken smoothly 
and the resultant free energy and entropy 
Eq.(\ref{e24}) and Eq.(\ref{e27}) agree with  
the known non-rotating results  
(in four dimension, see for example, \cite{kenmoku:}). 
 
\subsection{Area law 
in rotating black hole spacetime }
In this section, 
we apply the generalized Stephan-Boltzmann's law 
to the problem under consideration: 
the rotating black hole entropy.  
In order to estimate the entropy on the black hole horizon, 
we should impose the following horizon and temperature conditions
 on metric components $g^{tt}$, $g_{rr}$ and  
adopt the brick wall regularization scheme \cite{thooft:} 
to perform the real space integration 
of $V_{\rm opt}$ in Eq.(\ref{e27}).  
 
\begin{itemize}

\item[1.]{\large Horizon Condition}

We require simple zeros for the inverse metric components 
$1/g^{tt}$ and $1/g_{rr}$ at the black hole horizon $r_{H}$, 
which is the radius 
corresponding to the 
outer zero of these inverse metric components.  
We do not treat the extreme case of black holes in this paper, 
where the above inverse metric components have multi-zeros at 
the horizen. 
Then we can expand metric components  near horizon as 
\begin{eqnarray}
\frac{1}{g^{tt}}\simeq C_{t}(\theta)(r-r_{H})
\ \ , \ \  
\frac{1}{g_{rr}}\simeq C_{r}(\theta)(r-r_{H})
 \ ,  \label{e30}
\end{eqnarray}
where the coefficient functions 
$C_{t}(\theta), C_{r}(\theta)$ are defined by   
\begin{eqnarray}
C_{t}(\theta)
:=\left. \partial_{r}\frac{1}{g^{tt}}\right|_{r_{H}} \ , \ 
C_{r}(\theta)
:=\left. \partial_{r}\frac{1}{g_{rr}}\right|_{r_{H}}
\ . \label{e31}
\end{eqnarray}

\item[2.]{\large Temperature Condition}

As we are considering the thermodynamics for 
the scalar field around rotating black holes, 
it is necessary to define the temperature of this system.  
It is defined by the condition 
that no conical singularity is required in the Rindler space. 
This  gives the temperature as 
\begin{eqnarray}
\frac{2\pi}{\beta_{H}}=
\left. \frac{-\partial_{r} (1/g^{tt})}{2\sqrt{-g_{rr}/g^{tt}}}
\right|_{r_{H}}
=\frac{1}{2}\left({C_{t}(\theta)C_{r}(\theta)} \right)^{1/2}
\ . \label{e32}
\end{eqnarray}
As the temperature on the horizon should not 
depend on angle variables,  
we impose the condition that the product of 
the coefficient functions in Eq.(\ref{e32})
depends not on the zenithal angles 
but only on the horizon radius such that we can state    
\begin{eqnarray}
 C_{t}(\theta)C_{r}(\theta)
=\mbox{independent function on}\ \theta
\ . \label{e33}
\end{eqnarray}
We call this the horizon temperature condition, which 
will lead the area law for the entropy 
in the rotating black hole spacetime.   
\end{itemize}

Under these two conditions, 
we estimate the radial integration part in the optical volume 
Eq.(\ref{e26}) near the horizon and obtain 
\begin{eqnarray}
\int_{r_{H}+\epsilon}^{L}dr\, (-g^{tt})^{D/2-1/2}(g_{rr})^{1/2}
\simeq C_{t}(\theta)^{-D/2+1/2}
C_{r}(\theta)^{-1/2}\frac{\epsilon^{-D/2+1}}{D/2-1}
\ , \label{e34}
\end{eqnarray}
where $\epsilon$ and $L$ are the short distance and large 
distance regularization parameter respectively 
in the brick wall regularization scheme  
with their magnitudes 
restricted by the relation $0< \epsilon \ll r_{H}\ll L < \infty$. 
Eq.(\ref{e34}) is obtained by taking the large $L$ 
limit and is valid for $3\leq D$.   
Instead of $\epsilon$ we introduce 
a more physical cutoff parameter,  
the invariant cutoff parameter 
$\epsilon_{\rm inv}(\theta)$, as
\begin{eqnarray}
\epsilon_{\rm inv}(\theta)
:=\int_{r_{H}}^{r_{H}+\epsilon}dr\, (g_{rr})^{1/2}
\simeq 2{C_{r}(\theta)^{-1/2}{\epsilon}^{1/2}}
\ . \label{e35}
\end{eqnarray}
Combining together the radial integration Eq.(\ref{e34}) and 
the invariant cutoff Eq.(\ref{e35}) into the expression of 
optical volume Eq.(\ref{e26}), we obtain 
\begin{eqnarray}
V_{\rm opt}=
\left. \frac{2^{D-2}}{D/2-1}
\left(C_{t}C_{r}\right)^{-D/2+1/2} 
\int d\phi d\theta^{3} \cdots d\theta^{D-1}
\frac{(g_{\phi\phi}g_{33} \cdots g_{{D-1} {D-1}})^{1/2}}
{(\epsilon_{\rm inv}(\theta))^{D-2}}
\right|_{r_{H}}\ . \label{e36}
\end{eqnarray}
The optical volume is divergent as the cutoff parameter 
$\epsilon$ or $\epsilon_{\rm inv}(\theta)$ tends to zero, because 
of the large time dilation effect near the horizon.  
Combining the near horizen expressions of the temperature Eq.(\ref{e32}) 
with the condition Eq.(\ref{e33})
, the optical volume Eq.(\ref{e36}) and the unit sphere Eq.(\ref{e23}),  
we have a generalized form of the entropy: 
\begin{eqnarray}
S=\left. \frac{\zeta(D)D\Gamma(D/2-1)}{2^{D}\pi^{3D/2-1}}
\int d\phi d\theta^{3} \cdots d\theta^{D-1}
\frac{(g_{\phi\phi} g_{33} \cdots g_{{D-1} {D-1}})^{1/2}}
{(\epsilon_{\rm inv}(\theta))^{D-2}}
\right|_{r_{H}}\ .   \label{e37}
\end{eqnarray}
Note that the entropy expression Eq.(\ref{e37}) does not depend on   
the coefficient functions 
$C_{t}(\theta)$ and $ C_{r}(\theta)$. 
The invariant cutoff parameter $\epsilon_{\rm inv}(\theta)$
depends on zenithal angles $\theta$ and is 
included in the area integration in Eq.(\ref{e37}).  
In this sense, the obtained entropy could be thought of as  a zenithal  
angle dependent area law of the rotating black holes.

However, we change the idea 
from the $\theta$ independent cutoff parameter $\epsilon$ 
to the $\theta$ dependent cutoff parameter $\epsilon(\theta)$
defined through the integration 
\begin{eqnarray}
\epsilon_{\rm inv}:=\int_{r_{H}}^{r_{H}+\epsilon(\theta)}
dr \sqrt{g_{rr}}\ , \label{e37.1}
\end{eqnarray}
which is solved near the horizon as
\begin{eqnarray}
\epsilon(\theta)
= \frac{C_{r}(\theta)}{4}\epsilon_{\rm inv}^2 
\ , \label{e38}
\end{eqnarray}
where the invariant cutoff parameter $\epsilon_{\rm inv}$ 
is fixed to be constant. 
Using this constant invariant cutoff 
parameter, the entropy can be expressed as 
\begin{eqnarray}
S= \frac{\zeta(D)D\Gamma(D/2-1)}{2^{D}\pi^{3D/2-1}}
\frac{A}{({\epsilon_{\rm inv}})^{D-2}}
\ , \label{e39}
\end{eqnarray}
where $A$ is the surface area of the rotating black holes 
on the horizon 
\begin{eqnarray}
A:=\left. \int d\phi d\theta^{3} \cdots d\theta^{D-1}
(g_{\phi\phi} g_{33} \cdots g_{{D-1} {D-1}})^{1/2}
\right|_{r_{H}}\ .   \label{e40}
\end{eqnarray}
Eq.(\ref{e39}) exhibits the desired area law of the entropy 
in $D$ dimensional rotating black hole spacetime. 
Note that 
we have derived the result without using any explicit expression 
for the metric solutions but 
using only two conditions; Horizon Condition 
Eq.(\ref{e30}) and Temperature Condition Eq.(\ref{e33}).
Note also that we can take 
the smooth non-rotating limit of the generalized area law, 
which then reproduces the known expressions. 
This is the main result of this paper.

At this stage, we comment on the superradiant modes  which are 
characteristic of rotation black holes \cite{thorne:}.  
The horizon radius $r_{H}$ is defined to occur at the outer zero of 
the metric $1/g^{tt}$ and the ergosphere 
radius $r_{\rm ergo}$ is defined to occur at the 
outer zero of $g_{tt}$.  
The difference between these outer radii 
$r_{\rm ergo}- r_{H}\ (\geq 0)$ is due to 
the rotating geometry, and 
the region between them is the ergosphere. 
In the momentum picture, 
the situation corresponds to the energy restriction 
in Eq.(\ref{e13}): $E-m\Omega_{H}\geq 0$ .   
As azimuthal momentum of the scalar field $m$ can be negative 
as well as positive, the energy of the scalar fields can also 
be negative within the energy restriction, which correspond to 
the superradiant modes. 
We stress that the superradiant modes are included 
in our integration regions 
on radial coordinate $r\in [r_{H}+\epsilon, L]$ and 
energy $E \in [p_{\phi}\Omega_{H},\infty)$.

\section{Application}  

In this subsection, we apply the generalized area law of the entropy 
Eq.(\ref{e37}) or Eq.(\ref{e39}) 
to some special rotating black hole solutions. 

\subsection{BTZ black hole in (2+1) dimension}

This is the rotating black hole solution with 
a negative cosmological term $\Lambda(<0)$ 
 in (2+1) dimension.  The geometry is asymptotically 
anti-de Sitter spacetime. The unit of the gravitational 
constant is set to $G_{D=3}=1/8$. 
   
The line element of the BTZ black hole solution is given by
\begin{eqnarray}
ds^2=g_{tt}dt^2+g_{rr}dr^2 + g_{\phi\phi}d\phi^2+2g_{t\phi}dtd\phi 
\ , \label{f1}
\end{eqnarray}
with metric components 
\begin{eqnarray}
g_{tt}
&=&M-{r^2|\Lambda|}\ \ , \ \ 
g_{t\phi}=-\frac{J}{2}\ , \nonumber \\ 
g_{\phi\phi}&=& r^2  \ \ , \ \ 
{g_{rr}}=\left(-M+\frac{J^2}{4r^2}+{r^2|\Lambda|}\right)^{-1} \ 
\ ,  \label{f2}
\end{eqnarray}
where $M$ and $J$ are the mass and the angular 
momentum of the (2+1) dimensional black hole. 
Note that all metric components depend only on the radial coordinate $r$. 
The contravariant time component of the metric is given by  
\begin{eqnarray}
g^{tt}=-{g_{rr}} \ .  \label{f3}
\end{eqnarray}
The metric component $g^{tt}$ has two zeros at the radii given by
\begin{eqnarray}
r_{\pm}=
\left[\frac{ M}{2|\Lambda|}
\left(1\pm(1-\frac{|\Lambda|J^2}{M^2})^{1/2}\right)\right]^{1/2}
\ , \label{f4}
\end{eqnarray}
which are valid for the range:    
$0<M$ and $|J|\leq M/\sqrt{|\Lambda|}$.  
The event horizon is defined by $r_{H}=r_{+}$ 
and the radius of the ergosphere is given by the outer zero 
of $g_{tt}$ as, $r_{\rm erg}=\sqrt{M/|\Lambda|}$,  
which is larger than $r_{H}$.  
The off-diagonal component of the metric induces 
the rotating effect and the angular velocity on the horizon is given by 
\begin{eqnarray}
\Omega_{H}=\left. -\frac{g_{t\phi}}{g_{\phi\phi}}
\right|_{r_{H}}=\frac{J}{2r_{H}^2} \ . \label{f5}
\end{eqnarray}
Due to the metric relation Eq.(\ref{f3}) the temperature condition is 
satisfied and the temperature on the horizon is given by
\begin{eqnarray}
\frac{2\pi}{\beta_{H}}=|\Lambda|\frac{(r_{H}^2-r_{-}^2)}{r_{H}}
\ . \label{f6}
\end{eqnarray}
As the metric components are the function of $r$ only,    
the invariant cutoff parameter $\epsilon_{\rm inv}$ 
does not dependent on angle variable and is given, as a function 
of the cutoff parameter $\epsilon$, by   
\begin{eqnarray}
\epsilon_{\rm inv}=\int_{r_{H}}^{r_{H}+\epsilon}dr\, \sqrt{g_{rr}}
\simeq \left(
\frac{2r_{H}\epsilon}{|\Lambda|(r_{H}^2-r_{-}^2)}\right)^{1/2}
\ . \label{f7}
\end{eqnarray}
The area on the horizon in (2+1) dimension is the perimeter  
\begin{eqnarray}
 A=\left. \int_{0}^{2\pi}d\phi\, \sqrt{g_{\phi\phi}}\right|_{r_{H}}
= 2\pi r_{H}
\ ,  \label{f8}
\end{eqnarray}
and the entropy of the BTZ black hole 
from the generalized formula Eq.(\ref{e37}) or Eq.(\ref{e39}) is 
obtained as
\begin{eqnarray}
S_{D=3}=\frac{3\zeta(3)A}{(2\pi)^3\epsilon_{\rm inv}}
\ .  \label{f9} 
\end{eqnarray}
The entropy of the BTZ black hole Eq.(\ref{f9}) 
includes the rotation effect through the perimeter 
$A$ and the invariant cutoff $\epsilon_{\rm inv}$ 
via horizon $r_{H}=r_{+}$ in Eq.(\ref{f4}). 
The zero rotation limit can be taken smoothly.

\subsection{ Kerr-Newman black hole in (3+1) dimension}

Next we treat the Kerr-Newman black hole as an application to 
the (3+1) dimensional spacetime. 
This black hole solution is interesting because the metric components 
depend on zenithal angles $\theta$ as well as on the radial 
coordinate $r$.  The unit of the gravitational constant is set 
to $G_{D=4}=1$. 

The line element of the Kerr-Newman black hole is given by 
\begin{eqnarray}
ds^2=
g_{tt}dt^2+g_{rr}dr^2+g_{\phi\phi}d\phi^2+2g_{t\phi}dt d\phi
 +g_{\theta\theta} d\theta^2 \ , \label{g1}
\end{eqnarray}
with the metric components 
\begin{eqnarray}
g_{tt}&=&-\frac{\Delta-a^2\sin^2\theta}{\Sigma}\ \ , \ \ 
g_{t\phi}=-\frac{a\sin^2{\theta(r^2+a^2-\Delta)}}{\Sigma}\ \ , 
\nonumber \\
g_{\phi\phi}&=&\frac{(r^2+a^2)^2-\Delta a^2\sin^2{\theta}}
{\Sigma}\sin^2\theta\ \ , \ \ g_{\theta\theta}=\Sigma \ \ , 
\nonumber \\  
 g_{rr}&=&\frac{\Sigma}{\Delta}\ , \label{g2}
\end{eqnarray}
where 
\begin{eqnarray}
\Sigma (r,\theta):=r^2+a^2\cos^2\theta\ \ , \ \ 
\Delta(r):=r^2-2Mr+a^2+e^2 \ .\label{g3}
\end{eqnarray}
Here $M, a$ and $e$ denote the three hairs of the black holes: 
mass, angular momentum per unit mass 
and charge respectively.  
The time component of the inverse metric is 
\begin{eqnarray}
g^{tt}=-\frac{g_{\phi\phi}}{\Delta\sin^2\theta}
\ . \label{g4}
\end{eqnarray}
The inverse metric components $1/g^{tt}$ and $1/g_{rr}$ have two zeros 
for $\Delta =0$ at the radii 
\begin{eqnarray}
r_{\pm}=M\pm (M^2-a^2-e^2)^{1/2} \ \ , \ \ (a^2+e^2\leq M^2) 
\ , \label{g5}
\end{eqnarray}
among which horizon is $r_{H}=r_{+}$. 
The radius of the ergoregion is the outer zero of $g_{tt}$ and is 
given by $r_{\rm erg}=M+(M^2-a^2\cos^2\theta -e^2)^{1/2}$, which is 
larger than $r_{H}$.  \label{g6}
The angular velocity on the horizon is defined by the ratio of 
$g_{t\phi}$ and $g_{\phi\phi}$ and is given  
\begin{eqnarray}
\Omega_{H}=\frac{a}{r_{H}^2+a^2} \ . \label{g6.1}
\end{eqnarray}
The temperature condition Eq.(\ref{e33}) is satisfied due to 
the expressions of $g_{rr}$ in  Eq.(\ref{g2}) and $g^{tt}$ in 
Eq.(\ref{g4}) and the temperature on the horizon is given by  
\begin{eqnarray}
\frac{2\pi}{\beta_{H}}=\frac{r_{H}-r_{-}}{2(r_{H}^2+a^2)} \ . \label{g7}
\end{eqnarray}
The invariant cutoff parameter near the horizon is given by  
Eq.(\ref{e35}) as a function of  
the original cutoff parameter $\epsilon$, viz.,  
\begin{eqnarray}
\epsilon_{\rm inv}(\theta)=
2\sqrt{\frac{r_{H}^2+a^2\cos^2\theta}{r_{H}-r_{-}}}
\sqrt{\epsilon} \ . \label{g8}  
\end{eqnarray}
The entropy of the Kerr-Newman black hole is obtained from the 
general expression in Eq.(\ref{e37}) as 
\begin{eqnarray}
S=\left. 
\frac{1}{360 \pi}\int_{-\pi}^{\pi}d\phi \int_{0}^{\pi}d\theta 
\frac{(g_{\phi\phi}g_{\theta\theta})^{1/2}}
{(\epsilon_{\rm inv}(\theta))^2}\right|_{r_{H}} \ 
=\frac{A_{\rm rea}}{720 \pi}
 \int_{0}^{\pi}d\theta \frac{\sin \theta}
{(\epsilon_{\rm inv}(\theta))^2}
,  \label{g9}
\end{eqnarray}
in which the area of the Kerr-Newman black hole is given by 
\begin{eqnarray}
A=4\pi (r_{H}^2+a^2) \ .  \label{g10} 
\end{eqnarray}

Instead of using the constant cutoff parameter, 
the $\theta$ dependent cutoff parameter Eq.(\ref{e38}) 
is used, which, in this case, is     
\begin{eqnarray}
\epsilon(\theta) = \frac{(r_{H}-r_{-})^2}{4(r_{H}^2+a^2\cos^2\theta)}
\epsilon_{\rm inv}^2
\ , \label{g11}
\end{eqnarray}
and the generalized area law in  
the Kerr-Newman case is obtained  
\begin{eqnarray}
S=\frac{1}{360\pi}\frac{A}{{\epsilon_{\rm inv}}^2}
\ .  \label{g12}
\end{eqnarray}
The final form Eq.(\ref{g12})
is the same form as the non-rotating black hole cases, 
but the rotating effects are included in $A$ and implicitly 
in $\epsilon_{\rm inv}$ through $\theta$ dependent 
$\epsilon(\theta)$ paremeter. 

The results above show that the generalized area law works 
well enough giving known results. 
The non-rotating limit is straightforward.

\section{Another Derivation: 
 Euclidean Path Integral Method}

In this section, we try to derive the entropy formula 
of the scalar field in the rotating black hole spacetime 
by the Euclidean path integral method. This is a check 
of our result whether it depends on the calculation method or 
the approximation method.  

Using the Euclidean time $\tau:=it$, we first set the 
Euclidean $D$ dimensional polar coordinate as 
\begin{eqnarray}
x^{a}
=(x^1, x^2, x^3, \cdots , x^{D} )
=(r, \phi, \theta^{3}, \cdots , \theta^{D-1},\tau)
\ \ , \label{h0}
\end{eqnarray}
and rewrite the line element of Eq.(\ref{e2}) in 
Euclidean form as
\begin{eqnarray}
ds^2=g_{\tau\tau}d\tau^2 +g_{rr}dr^2 
+ g_{\phi\phi}d\phi^2+2g_{\tau\phi}d\tau d\phi 
+\sum_{a=3}^{D-1}g_{aa}(d\theta^{a})^2\ , \label{h1}
\end{eqnarray}
with $g_{\tau\tau}=-g_{tt}$ and $g_{\tau\phi}=-ig_{t\phi}$.  
Just as in the previous section, 
the metric components are assumed to be functions of $\tau$ and $\phi$ 
for the rotating spacetime and then two Killing vectors exist:
\begin{eqnarray}
\xi_{\tau}^{a}=(0,\cdots,0,1)\ , \ \ \xi_{\phi}^{a}=(0, 1, 0, \cdots, 0)\ 
\ , \label{h2}
\end{eqnarray}
which imply the energy and angular momentum conservations. 
The  Euclidean action $I_{\rm matter}$ and 
the Lagrangian density $\mathcal{L}$ 
of the scalar field of mass $\mu$
are written as 
\begin{eqnarray}
I_{\rm matter}&=&\int d^D x \sqrt{g_{E}}\ 
\mathcal{L} 
\ , \nonumber \\
\mathcal{L}&=&
-\sum_{a,b=1}^{D}\frac{1}{2}
g^{ab}\partial_{a}\Phi \partial_{b}\Phi
-\frac{1}{2}\mu^2\Phi^2  \ , \label{h3}
\end{eqnarray}
where $g_{E}$ denotes the determinant of the Euclidean metric.  
 The canonical momentum of the scalar field is given by 
\begin{eqnarray}
\Pi:=\frac{\partial \mathcal{L}}{\partial \dot{\Phi}}
=-g^{\tau\tau}\dot{\Phi}-g^{\tau\phi}\partial_{\phi}\Phi 
\ , \label{h4}
\end{eqnarray}
with the notation $\dot{\Phi}:=\partial \Phi/\partial \tau$. 
The quantization condition is given by 
\begin{eqnarray}
\left[\Phi(\tau,\boldmath{x}), \Pi(\tau, \boldmath{y}) \right]
=\frac{\delta^{D-1}(\boldmath{x}-\boldmath{y})}{\sqrt{g_{E}}}
\ . \label{h5}
\end{eqnarray}
The partition function with inverse temperature $\beta$ is 
\begin{eqnarray}
Z=Tr\left[{\rm e}^{-\beta(H-\Omega_{H}P_{\phi})}
\right]\ , \label{h6}
\end{eqnarray}
where $\Omega_{H}$ denotes the angular velocity of the black hole 
on the horizon.
The total energy $H$ and the azimuthal angular momentum $P_{\phi}$ of 
the scalar field are given by  
\begin{eqnarray}
H:&=&\int_{\Sigma} (\xi_{\tau})_{a}{\mathcal{T}}^{a \tau} d \Sigma_{\tau}
=\int d^{D-1}x \sqrt{g_{E}}\left(\Pi \dot{\Phi}-\mathcal{L}\right)
\ , \nonumber \\ 
P_{\phi}:&=&\int_{\Sigma}(\xi_{\phi})_{a}
{\mathcal T}^{a \tau} d \Sigma_{\tau} 
=- \int d^{D-1}x \sqrt{g_{E}}\, \Pi\, \partial_{\phi}{\Phi}
\ ,  \label{h7}
\end{eqnarray}
which involve the energy-momentum tensor 
\begin{eqnarray}
{\mathcal T}^{ab}:=-\frac{2}{\sqrt{g_{E}}} 
\frac{\delta I_{\rm matter}}{\delta g_{ab} }
\ , \label{h8}
\end{eqnarray}
and the Killing vectors $\xi$ as in Eq.(\ref{h2}). 
The exponent of the Boltzmann factor is the 
linear combination 
of these conserved quantities:  
\begin{eqnarray}
H-\Omega_{H} P_{\phi}  
= \int_{\Sigma} (\xi_{\tau}+\Omega_{H}\xi_{\phi})^{a}
T_{a \tau} d \Sigma^{\tau} 
=:  \int d^{D-1}x \sqrt{g_{E}} \mathcal{H}'  \ , 
\label{h9}
\end{eqnarray}
which should be non-negative according to the same argument 
as in Eq.(\ref{e13}). 
The newly defined Hamiltonian density $\mathcal{H}'$ is given by 

\begin{eqnarray}
&&{\mathcal{H}}' \nonumber \\
&=&
-\frac{1}{2g^{\tau\tau}}\Pi^2
+ \sum_{a=1}^{D-1}\frac{1}{2g_{aa}}{(\partial_{a} \Phi)^2}
+(\Omega_{H}+\frac{g_{\tau\phi}}{g_{\phi\phi}})\Pi\partial_{\phi}\Phi
 + \frac{\mu^2}{2}\Phi ^2   \nonumber \\
&\simeq& 
-\frac{1}{2g^{\tau\tau}}\Pi^2
 + \sum_{a=1}^{D-1}\frac{1}{2g_{aa}}(\partial_{a} \Phi)^2
 + \frac{\mu^2}{2}\Phi ^2   
\ , \label{h9.1}
\end{eqnarray}
 where sum runs $a=r, \phi, 3, \cdots , D-1$ . 
 The near horizon approximation has been used in Eq.(\ref{h9.1})
, corresponding to Eq.(\ref{e20}), such that
\begin{eqnarray}
-\frac{g_{\tau\phi}}{g_{\phi\phi}} \simeq\Omega_{H} 
\ \  \mbox{for}\ \ r\simeq r_{H} \ . \label{h10}
\end{eqnarray}
The contravariant $\tau$ component of the metric in Eq.(\ref{h9.1}) is 
$g^{\tau\tau}=g_{\phi\phi}/(g_{\tau\tau}g_{\phi\phi}-g_{\tau\phi}^2)$, 
which does not equal to $1/g_{\tau\tau}$ for rotating case. 
Turning back to the calculation of the partition function, 
we express it in the Euclidean path integral form 
\begin{eqnarray}
Z=\int [D\Phi\,D\Pi\,{g_{E}^{1/2}}] \,
\exp{\left(\int d^{D}x \, \sqrt{g_{E}}\,
(\Pi\dot{\Phi}-\mathcal{H}')\right)} 
\ . \label{h10.1}
\end{eqnarray}
We perform the momentum integration and obtain 
after the integration by parts as
\begin{eqnarray}
Z=\int [D \Phi g_{E}^{1/4} (g^{\tau\tau})^{1/2}]
\exp{(
-\int d^{D}x\, \frac{\sqrt{g_{E}} g^{\tau\tau} }{2}
\Phi\, \bar{K}\, \Phi}) \ , \label{h11}
\end{eqnarray}
where $\bar{K}$ denotes the kernel in the optical space 
given by 
\begin{eqnarray}
\bar{K}
&:=& - \partial_{\tau}^2 -
\frac{1}{g^{\tau\tau}}(\frac{1}{\sqrt{g_{E}}}
\sum_{a=2}^{D}\partial_{a}\frac{\sqrt{g_{E}}}{g_{aa}}\partial_{a} 
-\mu^2) 
 \ .  \label{h12}
\end{eqnarray}
After the Gaussian integration with respect to $\Phi$, 
we obtain the free energy using the heat kernel representation 
\cite{birrell:} as
\begin{eqnarray}
\beta F 
&=& \frac{1}{2}\ln \det \bar{K} 
= \frac{1}{2} \mbox{Tr} \ln \bar{K} \ \nonumber \\ 
&=&- \frac{1}{2} \mbox{Tr} 
\int_{0}^{\infty}\frac{ds}{s}\exp{(-s\bar{K})}
 \ . \label{h13}
\end{eqnarray}
The trace of the heat kernel is divided into two parts: 
the Euclidean time part and the space part.  
The Euclidean time part is evaluated by using the one dimensional  
eigenfunction of $-i\partial_{\tau}$ as 
\begin{eqnarray}
{\rm Tr} \ \exp{(s\, \partial _{\tau}^2)} &=& 
\int_{0}^{\beta} d\tau \sum_{\ell=-\infty}^{\infty}\frac{1}{\beta}
\exp{(-s(\frac{2\pi \ell}{\beta})^2)} \nonumber \\
&=& \sum_{n=-\infty}^{\infty}\frac{\beta}{(4\pi s)^{1/2}}
\exp{(-\frac{\beta^2 n^2}{4s})} \  . \label{h14}
\end{eqnarray}
The Poisson's summation formula is used in the second equality 
in order to transform the low temperature expansion around    
$s \rightarrow 0$ to the high temperature expansion around   
$1/s \rightarrow 0$.  
The space part of the trace is evaluated by the asymptotic expansion 
method 
\begin{eqnarray}
&&\mbox{Tr}\, \exp{(\frac{s}{g^{\tau\tau}} 
(\frac{\partial_{\phi}^2}{g_{\phi\phi}}
+\sum_{a=1}^{D-1}\frac{\partial_{i}^2}{g_{ii}}
+\frac{\partial_{r}^2}{g_{rr}} -\mu^2))} \nonumber \\
&=&\frac{1}{(4\pi s)^{(D-1)/2}}
\sum_{k=0}^{\infty}\bar{B}_{k}{(-s)^k}
\exp{(-\frac{s\mu^2}{g^{\tau\tau}})} \nonumber\\
&\simeq& \frac{1}{(4\pi s)^{(D-1)/2}}
\bar{B}_{0}\exp{(-\frac{s\mu^2}{g^{\tau\tau}})}
\ , \label{h15}
\end{eqnarray}
where $\bar{B}_{k}$'s are the coefficient functions   
of the asymptotic expansion and only the lowest contribution  
$\bar{B}_{0}$ is taken account, which is 
the integration function in the optical space, given by 
\begin{eqnarray}
\bar{B}_{0}= 
\int d^{D-1}x (g^{\tau\tau})^{(D-1)/2}
(g_{rr}g_{\phi\phi}g_{33} \cdots g_{{D-1}{D-1}})^{1/2}
\ . \label{h16}
\end{eqnarray}
In the asymptotic expansion, 
the derivative terms of the metric are considered to be small and 
neglected in the lowest contribution. \footnote[3]
{Note that 
the lowest order contribution of the asymptotic expansion 
in the trace calculation Eq.(\ref{h15}) can also be derived 
by using the semiclassical momentum eigenfunction ( Eq.(\ref{e6})) 
with normalization factor  and by the Gaussian integration over momenta.  
This confirms the result of the lowest contribution from another side.} 
The free energy is expressed by multiplying the two trace parts and 
we obtain 
\begin{eqnarray}
F=&-&\int_{0}^{\infty}\frac{ds}{s}\frac{1}{(4\pi s)^{D/2}}
\sum_{n=1}^{\infty}
\exp{(-\frac{\beta^2n^2}{4s})} \bar{B}_{0} 
\exp{(-\frac{\mu^2s}{g^{\tau\tau}})} \nonumber\\
=
&-&\frac{1}{\beta^D \pi^{D/2}}
\int_{0}^{\infty}\frac{dt}{t}t^{D/2}{\rm e}^{-t}
\sum_{n=1}^{\infty}\frac{1}{n^D} \bar{B}_{0}
\exp{(-\frac{\mu^2\beta^2n^2}{4tg^{\tau\tau}})} \ , 
\label{h17}
\end{eqnarray}
where the integration variable in second equality 
has been changed according to  
\begin{eqnarray}
t=\frac{\beta^2n^2}{4s}\ . 
\label{h18}
\end{eqnarray}
The temperature independent term $(n=0)$ 
is subtracted in the sum of Eq.(\ref{h17}). 
The massless case $(\mu=0)$ may be important and 
in this case we have a compact 
expression for the free energy 
\begin{eqnarray}
F=-\frac{\zeta(D)\Gamma(D/2)}{\pi^{D/2}\beta^{D}}\ V_{\rm opt}
\ , \label{h19}
\end{eqnarray}
where we have used the equality relation 
$\bar{B}_{0}\times 1=V_{\rm opt}$ 
for the massless case,
 where $V_{\rm opt}$ is defined in Eq.(\ref{e26}) in the previous section. 
The massless expression for the free energy obtained 
by the Euclidean path integral method in Eq.(\ref{h19})  
completely coincides with that obtained by the semiclassical method 
in Eq.(\ref{e24}) with the use of the expression of $v_{\rm unit}$ 
in Eq.(\ref{e23}). 
Non-rotating limit of the free energy 
can be taken smoothly and agrees with 
the previous result \cite{dealwis:}. 

\vspace{3mm}
\noindent
{\bf \large Scalar field mass contribution}\\
So far we have studied the massless scalar case using two methods, 
semiclassical and the Euclidean path integral method, and 
find that the thermodynamic quantities are the same. 
How about the mass contribution?
In general, it is not so easy to estimate this.  
Only the small mass contribution to the free energy 
 can be studied by the perturbation method in four dimension.    
The zeroth order free energy in $D=4$ is given, from Eq.(\ref{e24}) or 
Eq.(\ref{h19}), by 
\begin{eqnarray}
F_{0}^{\rm (4-dim.)}=-\frac{\pi^2}{90 \beta^4}\int dr d\phi d\theta 
(g^{\tau\tau})^{3/2}(g_{rr}g_{\phi\phi}g_{\theta\theta})^{1/2} 
\ , 
\label{h20}
\end{eqnarray}
which is the same for two methods, of course. 
In right hand side, the space integration part is the 
four dimensional optical volume itself. 
The next order with respect to the scalar field mass contribution 
to the free energy is calculated from 
Eq.(\ref{e22}) or Eq.(\ref{h17}) as 
\begin{eqnarray}
F_{1}^{\rm (4-dim.)}=\frac{\mu^2}{24\beta^2}\int dr d\phi d\theta 
(g^{\tau\tau}g_{rr}g_{\phi\phi}g_{\theta\theta})^{1/2} 
\ , \label{h20.1}
\end{eqnarray}
which again happens to coincide with for two methods. 
It is worthwhile noting that 
the next order mass contribution does not show 
the Stephan-Boltzmann's law.   
The short distance singular behavior is the logarithmic 
divergent with respect to the cutoff parameter $\epsilon$ 
and is less singular than the first order contribution, 
and does not show the area law as well.  

Here we comment on the allowed range of enery. This  
is determined by two restrictions: the threshold energy from (\ref{e8}),  
$E \geq -\mu/\sqrt{g^{tt}}$, and 
the rotating effect on the energy in (\ref{e13}),  
$E \geq m\Omega_{H}$.   
The threshold effect to the free energy 
is shown to be negligible for $D\geq 3$ 
in the small mass case in the semiclassical method. 
The situation is similar in the Euclidean path integral method. 
Further we can show that 
the small mass contribution to the 
free energy in arbitraly $D$ dimension is also the same in both methods: 
\begin{eqnarray}
F_{1}^{(D\rm -dim.)}=\frac{\mu^2}{4\pi^{D/2}\beta_{H}^{D-2}}
\Gamma(\frac{D}{2}-1)\zeta(D-2)\, 
\bar{B_{0}}\, \frac{1}{g^{\tau\tau}} \ , \label{m1} 
\end{eqnarray}
which is less singular than zeroth order free energy,  
where $\bar{B_{0}}$ is given in (\ref{h16}) and 
does not show the area law as well.   

\section{Summary and Discussions}

We have studied the statistical mechanics of the scalar field in 
$D$ dimensional rotating spacetime. We have imposed the following 
physically admissible minimal set of conditions on the 
one parameter black hole metric.

\begin{itemize}
\item[a1.] Metric components are assumed not to be functions of $t$ and 
$\phi$ which ensure the energy and angular momentum conservations.

\item[a2.] Off-diagonal component of the metric, viz., $g_{t\phi }$ is
assumed to exist which indicates rotation with the angular velocity 
$-g_{t\phi }/g_{\phi \phi }$.

\item[a3.] Horizon condition is imposed on the inverse of metric
components, that is, $1/g^{tt}$ and $1/g_{rr}$ to have simple zeros at 
$r_{H} $ respectively.

\item[a4.] Temperature condition is also imposed on $\partial
_{r}(1/g^{tt})\times \partial _{r}(1/g_{rr})$ at $r_{H}$ to be $\theta $
independent. This condition leads the $\theta $ independent horizon
temperature.
\end{itemize}

Further, we have used some approximations:

\begin{itemize}
\item[b1.] Semiclassical method is adopted for the calculation.

\item[b2.] Near horizon approximation is adopted. The implication of this
approximation is that the angular velocity of the scalar field near the
horizon is the same as that of the black hole on the horizon.

\item[b3.] Mass contribution of the scalar field is neglected because the
high temperature limit is considered mainly.
\end{itemize}

Under these conditions and approximations, we derived the generalized
Stefan-Boltzmann's law and the generalized area law of the rotating black
holes in arbitrary $D$ dimensional spacetime. One of the key point is the
introduction of the zenithal angle dependent cutoff parameter 
$\epsilon (\theta )$, which leads to the constant invariant cutoff parameter 
$\epsilon _{{\rm inv}}$ and the generalized area law in a compact form for 
rotating black holes. The generalized area law is applied to the BTZ black
hole in (2+1) dimension and the Kerr-Newman black hole in (3+1) dimension.
Non-rotating limit of these thermal quantities can be taken smoothly and
they straightforwardly reproduce the known results.

We also adopted the Euclidean path integral method for the quantized scalar
field with the asymptotic expansion and derived results for the free energy
in rotating black hole spacetime that are consistent with the semiclassical
method. Small scalar field mass contribution has been assessed both in the
semiclassical and in the Euclidean methods and the same result followed. The
mass contribution to the entropy is less singular than the massless case and
doesn't show the area law.

There are other methods to study the statistical mechanics of the scalar
field in black hole spacetime. Among them, the quasi-normal mode approach 
\cite{quasi:} and the Green's function approach are
interesting because more rigorous treatments than the present methods could
be possible in those approaches, especially in the (2+1) dimension 
\cite{ichinose:,birmingham:}. Another interesting problem would be to study the
thermodynamics of the multi-parameter 
rotating black hole with the cosmological term 
in multi-dimension \cite{h:,g:,y:} because the existence of the
cosmological term is suggested in the observations and 
the recent developement in the view of AdS/CFT correspondence 
is interesting. These are problems we intend to work 
in the future \cite{k:}.



\begin{thebibliography}{99}
\bibitem{k:}
M. Kenmoku, Many parameter cases of rotating black holes 
will be studied in a separate paper.
\end{thebibliography}
\end{document}